 \documentclass[review,1p,times]{elsarticle}

\usepackage{graphicx}
\usepackage{amssymb}
\usepackage{amsmath}
\usepackage{xcolor}
\usepackage{xurl}
\usepackage{dirtytalk}
\usepackage{caption}
\usepackage{lineno} 

\journal{}

\begin{document}

\begin{frontmatter}


\title{Modeling the sea-surface \textit{p}CO\textsubscript2 of the central Bay of Bengal region using machine learning algorithms}



\author{ A.P Joshi\corref{label1}\textsuperscript{1*}, V. Kumar\textsuperscript{1}, and H.V Warrior\corref{cor2}\textsuperscript{1}}

\address{\textsuperscript{1}Department of Ocean Engineering and Naval Architecture, IIT Kharagpur, Kharagpur-721302, West Bengal, India}

\cortext[label1]{apurvajoshi@iitkgp.ac.in}

\begin{abstract}
The present study explores the capabilities of advanced machine learning algorithms in predicting the sea-surface \textit{p}CO\textsubscript2 in the open oceans of the Bay of Bengal (BoB). We collect the available observations (outside EEZ) from the cruise tracks and the mooring stations. Due to the paucity of data in the BoB, we attempt to predict \textit{p}CO\textsubscript2 based on the Sea Surface Temperature (SST) and the Sea Surface Salinity (SSS). Comparing the MLR, the ANN, and the XGBoost algorithm against a common dataset reveals that the XGBoost performs the best for predicting the sea-surface \textit{p}CO\textsubscript2 in the BoB. Using the satellite-derived SST and SSS, we predict the sea-surface \textit{p}CO\textsubscript2 using the XGBoost model and compare the same with the in-situ observations from RAMA buoy. The model performs satisfactorily, having a correlation of 0.75 and the RMSE of $\pm$ 12.23 $\mu$atm. Further using this model, we emulate the monthly variations in the sea-surface \textit{p}CO\textsubscript2 for the central BoB between 2010-2019. Using the satellite data, we show that the central BoB is warming at a rate of 0.0175 per year, whereas the SSS decreases with a rate of -0.0088 per year. The modeled \textit{p}CO\textsubscript2 shows a declination at a rate of -0.4852 $\mu$atm per year. We perform sensitivity experiments to find that the variations in SST and SSS contribute $\approx$ 41\% and $\approx$ 37\% to the declining trends of the \textit{p}CO\textsubscript2 for the last decade. Seasonal analysis shows that the pre-monsoon season has the highest rate of decrease of the sea-surface \textit{p}CO\textsubscript2.

\end{abstract}


\end{frontmatter}


\section{Introduction}
\label{S:1}

The escalation of the anthropogenic activities post the pre-industrial era has resulted in the proliferation of the atmospheric CO\textsubscript2. About 30\% of the man-made atmospheric CO\textsubscript2 is absorbed by the ocean \citep{sabine2004oceanic}, this highlights the importance of the ocean in regulating the atmospheric CO\textsubscript2. In the previous decade, the rate of CO\textsubscript2 absorption by the ocean is reported to be \ensuremath{\approx 2.5 \pm 0.6} GtC per year \citep{friedlingstein2020global}. For the year 2020, the ocean absorbs atmospheric CO\textsubscript2 at a rate of \ensuremath{\approx 3.0 \pm 0.4} GtC per year \citep{friedlingstein2021global}. The continuous rise in the atmospheric CO\textsubscript2 absorption rate by the ocean increases its sink strength \citep{regnier2013anthropogenic,bauer2013changing}. The sink strength of the global oceans is reduced by \ensuremath{\approx} 0.2 PgC per year \citep{laruelle2014regionalized}. On the contrary, more recent studies suggest an increasing sink strength of the continental shelves \citep{laruelle2018continental}. The South China Sea, which was once reported to be a weak source \citep{dai2013some,zhai2013seasonal} has changed into a sink of the atmospheric CO\textsubscript2 \citep{li2020partial}. While there is an increasing need to understand the dynamics and study the trends of the sea-surface \textit{p}CO\textsubscript2, the scarcity of measured data looms as a major impediment for such studies \citep{borges2005we,anderson2005plankton,anderson2010progress}.

The Bay of Bengal (BoB) is encircled by the landmasses of Southeast Asia and India. The idiosyncrasy of the BoB is attributed to large freshwater influx (through local precipitation and river discharge) \citep{unesco1969discharge} and the seasonal reversing of the currents \citep{shetye1996hydrography}. Due to the high freshwater influx, the BoB becomes a highly stratified ocean, which influences the physical dynamics (formation of Barrier Layer, freshwater plume spreading, and low vertical mixing). These altering physical dynamics influence the sea-surface \textit{p}CO\textsubscript2 distribution and this region's source and sink characteristics \citep{joshi2021influence}. The nutrients from the rivers influence the sea-surface \textit{p}CO\textsubscript2 near the coast, but the influence decreases as we move away from the coast due to the stratification \citep{sarma2021influence}. During March-April of 1991, the increased biological productivity due to the river-induced nutrients lowering the sea-surface \textit{p}CO\textsubscript2 creating a sink of the atmospheric CO\textsubscript2.   

The rivers of the north (Ganges \ensuremath{\approx} 500 \ensuremath{\mu atm}) have lower \textit{p}CO\textsubscript2 than the peninsular rivers (5000-17000 \ensuremath{\mu atm}), which influences the sea-surface \textit{p}CO\textsubscript2 in the coastal regions. During the South West monsoon (SWM) the north-western coast acts as a sink to the atmospheric CO\textsubscript2 (205 \ensuremath{\pm} 24 \ensuremath{\mu atm}), whereas the southwestern coast is a source (505 \ensuremath{\pm} 77 \ensuremath{\mu atm}). Since the sea-surface \textit{p}CO\textsubscript2 decreases due to the low saline waters from the northern rivers, the north BoB has lower sea-surface \textit{p}CO\textsubscript2 than the south BoB \citep{sarma2012sources,joshi2020configuration,joshi2021influence}. Also, the nutrients brought down by the northern rivers increases the biological productivity, which assists in the lowering of the sea-surface \textit{p}CO\textsubscript2 \citep{sarma2012sources}. 

The hydrographic characteristics of the BoB are governed by the seasonal reversing currents, known as the East Indian Coastal Current (EICC) or the Western Boundary Currents (WBC). From February to May, the EICC moves northward, bringing in high saline waters from the south to the coastal regions. This weakens the stratification and aids in the coastal upwelling. The upwelled waters from March-August brings the high subsurface \textit{p}CO\textsubscript2 (\ensuremath{\approx} 650 \ensuremath{\mu atm}) and Dissolved Inorganic Carbon (DIC) to the surface \citep{sarma2018east,joshi2020configuration}. From Oct-Dec the EICC moves in the south, increasing the stratification in the coastal regions as it transports low saline waters from the north. The strong stratification further aids in the reduction of the sea-surface \textit{p}CO\textsubscript2 (\ensuremath{\approx} 320 \ensuremath{\mu atm}) and increases the coastal sea-surface acidic levels (8.03 in Mar-Aug to 8.22 in Oct-Dec). The biological and thermal mechanisms rule the coastal sea-surface \textit{p}CO\textsubscript2 dynamics \citep{sarma2018east}.

The circulation pattern of the BoB is eddy-dominated. These eddies notably regulate the sea-surface \textit{p}CO\textsubscript2 by controlling the vertical mixing. The cyclonic eddies enhance vertical mixing, which increases the sea-surface \textit{p}CO\textsubscript2 and chlorophyll-a (chl-a) in the cyclonic eddy zone \citep{sarma2019impact}. The mixing and the biological mechanism influences the variability of the sea-surface \textit{p}CO\textsubscript2 in the cyclonic eddy zones, while mixing and thermal effects impact the variability of \textit{p}CO\textsubscript2 in the anti-cyclonic regions \citep{sarma2019impact}. The biological process has significant influence over the sea-surface \textit{p}CO\textsubscript2 in the coastal regions, but the influences consistently decreases towards the central open oceans \citep{sarma2021influence}. Similarly, the effect of atmospheric dust is localized, hence affecting only the coastal regions. The values of \textit{p}CO\textsubscript2 increases due to the atmospheric pollutants, in the winter and the spring seasons, along the western coast and the head bay region. In the strong upwelling zone, near the Srilankan dome, the sea-surface \textit{p}CO\textsubscript2 is influenced by the biological pump (net decrement of 21 \ensuremath{\mu atm}) \citep{chakraborty2018dominant}. 

When compared the coastal sea-surface \textit{p}CO\textsubscript2 observation between 1991 and 2011, it reveals that the coastal waters less saline and had lower \textit{p}CO\textsubscript2 values in 1991 \citep{sarma2015observed}. The rate of the increment of the sea-surface \textit{p}CO\textsubscript2 in the southwestern coast is \ensuremath{\approx} 1.5 \ensuremath{\mu atm} per year, whereas the rate is 3-5 times higher in the north-western coast (6.7 \ensuremath{\mu atm} per year). The increase fluxes in the nitrogen and sulphate aerosols in the north-western region could be the reason for the highers rates \citep{sarma2015observed}. Using the multi-linear regression model, \cite{sridevi2021role} shows that the sea-surface \textit{p}CO\textsubscript2 decreases overall in the BoB over a period from 1998-2015, except for the head bay region. The post-monsoon season displays the highest positive increment (+2.4 \ensuremath{\mu atm} per year) in the head bay \citep{sridevi2021role}. Due to the global warming the discharge rate of the Himalayan rivers is constantly increasing due to ice-melting. The SSS is a major driver controlling the spatiotemporal variability of the sea-surface \textit{p}CO\textsubscript2 in the northern BoB \citep{chakraborty2021seasonal,sridevi2021role}. 

In the past, researchers attempt to understand the dynamics of the sea-surface \textit{p}CO\textsubscript2 over a long period or during a cyclonic period by using observations or regression models \citep{dixit2019net,sridevi2021role,mohanty2022surface}. The physical-biogeochemical model coupling method \citep{chakraborty2018dominant,joshi2020configuration,chakraborty2021seasonal,joshi2021influence} are also popularly used, but the complexity (both time and numerical) of these models are huge. With the available observations increasing, from real-time buoy, and possible implementation of the proposed locations for collecting data by \cite{valsala2021observing}, the use of more complex machine learning based algorithms could be helpful. In the present study, we attempt to demonstrate the capability of machine learning algorithms and artificial neural network (ANN) in modeling the sea-surface \textit{p}CO\textsubscript2. We compare the performance to understand which algorithm performs the best. Then using the best algorithm, we try to emulate the sea-surface \textit{p}CO\textsubscript2 using satellite data. Further, we show the temporal variation in the \textit{p}CO\textsubscript2 in the central BoB and analyze the seasonal as well as interannual trends over the past decade (2010-2019).

\section{Data and Methodology}
\label{s:2}
\subsection{Data Collection}
\label{s:2.1}

\begin{figure}[ht!]
\centering
\includegraphics[height=10cm,width=11cm]{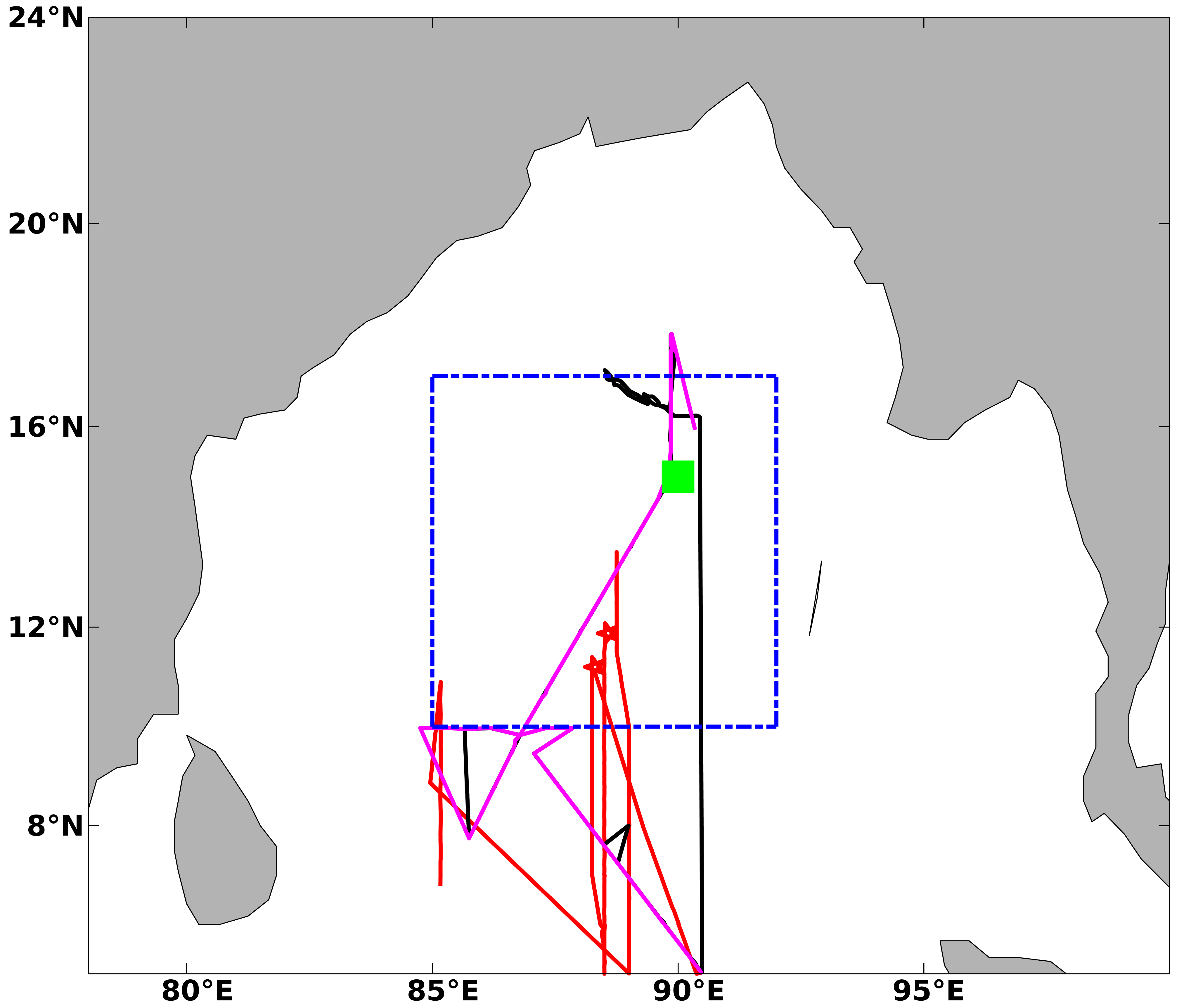}
\caption{The broken blue lines enclose the present study region. The green square shows the Rama Buoy location (15\textdegree{} N, 90\textdegree{} E). The red line represent the SOCAT track for the year 1999, the black line represents the SOCAT track for the year 2007, and the pink line represents SOCAT track for 2016.}
\label{fig:1}
\end{figure}

Fig.\ref{fig:1} shows the BoB basin, along with the study domain (blue broken lines) ranging from 10\textdegree{} N - 17\textdegree{} N in the longitudinal direction and from 85\textdegree{} E - 92\textdegree{} E in the latitudinal direction. The study domain represents the central open oceans of the BoB. The region being away from the coast is expected to have a minimum dependency on the biological productivity and the atmospheric dust, as these mechanisms are limited to the coastal zones \citep{sarma2021influence}. The green square in Fig.\ref{fig:1} represents the RAMA buoy location. The solid lines mark the tracks of the SOCAT database. The red line represents the track from 1999, the black line from 2007, and the pink line from 2016.  

\subsubsection{RAMA Buoy}
\label{s:2.1.1}

\begin{table}[ht!] 
\centering
\begin{tabular}{|c|c|}
\hline
Location & 15\textdegree{} N 90\textdegree{} E  \\
\hline
Type & Timeseries data  \\
\hline
Temporal Resolution & 3 hourly data  \\
\hline
 & 24 Nov 2013 to 6 Dec 2014 (Deployment 1) \\
Duration & 6 Dec 2014 to 19 Jun 2015 (Deployment 2) \\
 & 6 Mar 2016 to 9 Jan 2017 (Deployment 3) \\
 & 11 Jan 2017 to 20 Nov 2018 (Deployment 4) \\
 \hline
 Parameters & Seawater \textit{p}CO\textsubscript2, pH, SST, SSS \\
 \hline 
 Investigator Institution & Pacific Marine Environmental Laboratory \\
 \hline
\end{tabular}
\caption{RAMA mooring data details (adopted from \citep{joshi2020configuration,joshi2021influence}).}
\label{tab:1}
\end{table}

The RAMA buoy at 15\textdegree{} N, 90\textdegree{} E, also known as the Bay of Bengal Ocean Acidification (BOBOA) moored buoy is the only source of standardized and sustainable sea-surface \textit{p}CO\textsubscript2 and pH data available for a continuous period in the BoB region. The mooring is a part of Research Moored Array for African-Asian-Australian Monsoon Analysis and Prediction (RAMA) network. As BoB experiences major cyclonic events on a fairly regular basis, the continuous deployment of the buoy is difficult. Hence as shown in Table \ref{tab:1}, the buoy has four active periods. Apart from the measured parameters shown in Table \ref{tab:1}, the buoy also records wind-speed, currents, precipitation, and density. \cite{sutton2014high} provides an elaborate explanation on the methods adopted for calculating different carbonate parameters using this mooring. The data is available at \url{https://www.nodc.noaa.gov/ocads/oceans/Moorings/BOBOA.html}.

\subsubsection{SOCAT data}
\label{s:2.1.2}

The Surface Ocean CO\textsubscript2 Atlas database, SOCATv2019 \citep{becker2020northern} is the cruise observed sea-surface \textit{f}CO\textsubscript2 (the fugacity of the CO\textsubscript2) database. We took the available observations for the period 1999-2019, among which only three years provided the cruise data (1999, 2007, and 2016). The spatial coverage of cruise tracks is limited to the central and southern BoB. The SOCAT defines continental margin a to be 400 km hence we select points outside the 400 km shelf mark. The \textit{f}CO\textsubscript2 is converted to \textit{p}CO\textsubscript2 using the MATLAB CO2SYS program \citep{van2011matlab}. The SOCAT data is obtained from \url{https://www.socat.info/index.php/data-access/}.

\subsubsection{Satellite data}
\label{s:2.1.3}

In this study, we use the satellite data to reproduce the sea-surface \textit{p}CO\textsubscript2 using the best machine learning algorithm, and to understand the trend of \textit{p}CO\textsubscript2 in the central BoB. The source of the Sea Surface Temperature (SST) and the Sea Surface Salinity (SSS) are as follows:

\begin{itemize}
    \item \textbf{Sea Surface Temperature (SST)} :
    We obtain the SST from a Group for High Resolution Sea Surface Temperature (GHRSST) \citep{donlon2007global}. The data is an accumulation of various data (re-processed ATSR data, AVHRR Pathfinder data, and in-situ data from ICOADS), which enables a gap-free daily data with 0.05\textdegree{} $\times$ 0.05\textdegree{} resolution. The data is made available by GHRSST, Met office, and CMEMS. We collect the data from the Asia-Pacific Data-Research Centre (APDRC) (\url{http://apdrc.soest.hawaii.edu/las/v6/dataset?catitem=12590}).
    
    \item \textbf{Sea Surface Salinity (SSS)} :
    The SSS, especially for the BoB region, is complex to capture due to the cloud coverage over this region. Very few satellite products perform excellently with the observations for the BoB. \cite{akhil2020bay} shows that the merged dataset, which combines SMOS, Aquarius, and SMAP satellite data, performs exceedingly well in the BoB region. This merged dataset is a part of the Climate Change Initiative (CCI) of the European Space Agency (ESA) \citep{boutinesa}. The data has a spatial resolution of 0.25\textdegree{} $\times$ 0.25\textdegree{}, and is available on a bi-weekly scale. The ESA-CCI SSS is available from 2010 to 2019. We retrieve this data from \url{https://catalogue.ceda.ac.uk/uuid/fad2e982a59d44788eda09e3c67ed7d5}.
\end{itemize}

Before using the satellite data in this study, we convert them into monthly temporal resolution. Since the horizontal resolution of GHRSST is finer than the ESA-CCI data, so we interpolate the GHRSST to the horizontal resolution of the ESA-CCI SSS data, using \say{nearest-neighbour} interpolation technique. 

\subsection{Methods}
\label{s:2.2}

\subsubsection{Data visualization}
\label{s:2.2.1}

Since the observational data of the sea-surface \textit{p}CO\textsubscript2 for the BoB is low, before applying the machine learning algorithms, it is essential to understand the amount of data available and its distribution. For this study, we choose SST and SSS to be our independent variables and the \textit{p}CO\textsubscript2 as our dependent variable. We understand that there are several other factors that may affect the sea-surface \textit{p}CO\textsubscript2, but with the focus of the open central ocean of the BoB, we can safely assume that the factors apart from SST and SSS will have a more negligible effect. Another reason for not selecting more independent variables is that mostly these variables would be in gridded form, and since the cruise data are only a handful, we choose to take the non-gridded form of the data. If the gridded set of the cruise data is selected, it reduces the number of data points drastically, making applying machine learning algorithms more difficult (risk of over-fitting). 

\begin{figure}[ht!]
\centering
\includegraphics[height=8cm,width=10cm]{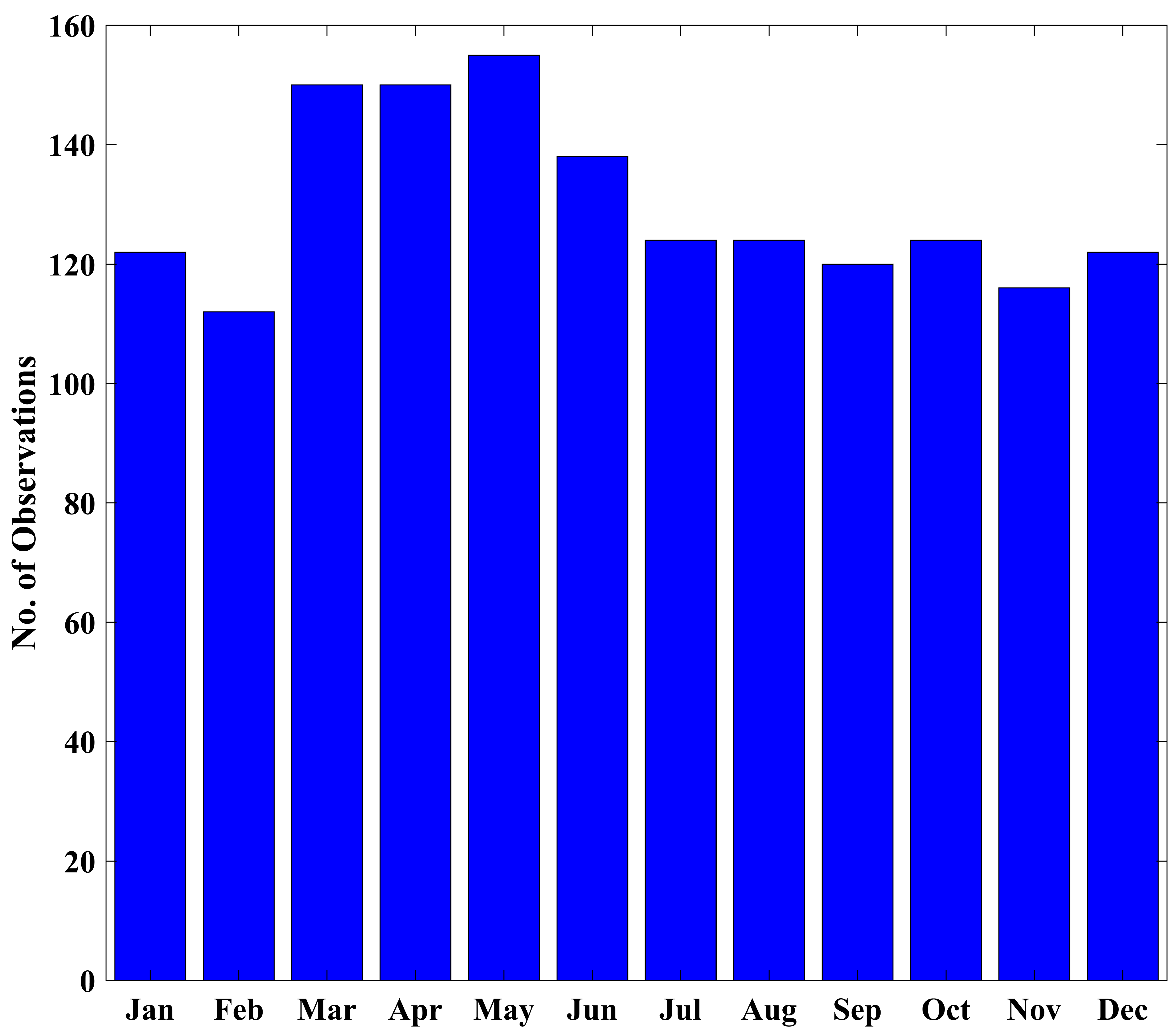}
\caption{Data available for each month from the RAMA buoy.}
\label{fig:2}
\end{figure}

\begin{figure}[ht!]
\centering
\includegraphics[height=8cm,width=10cm]{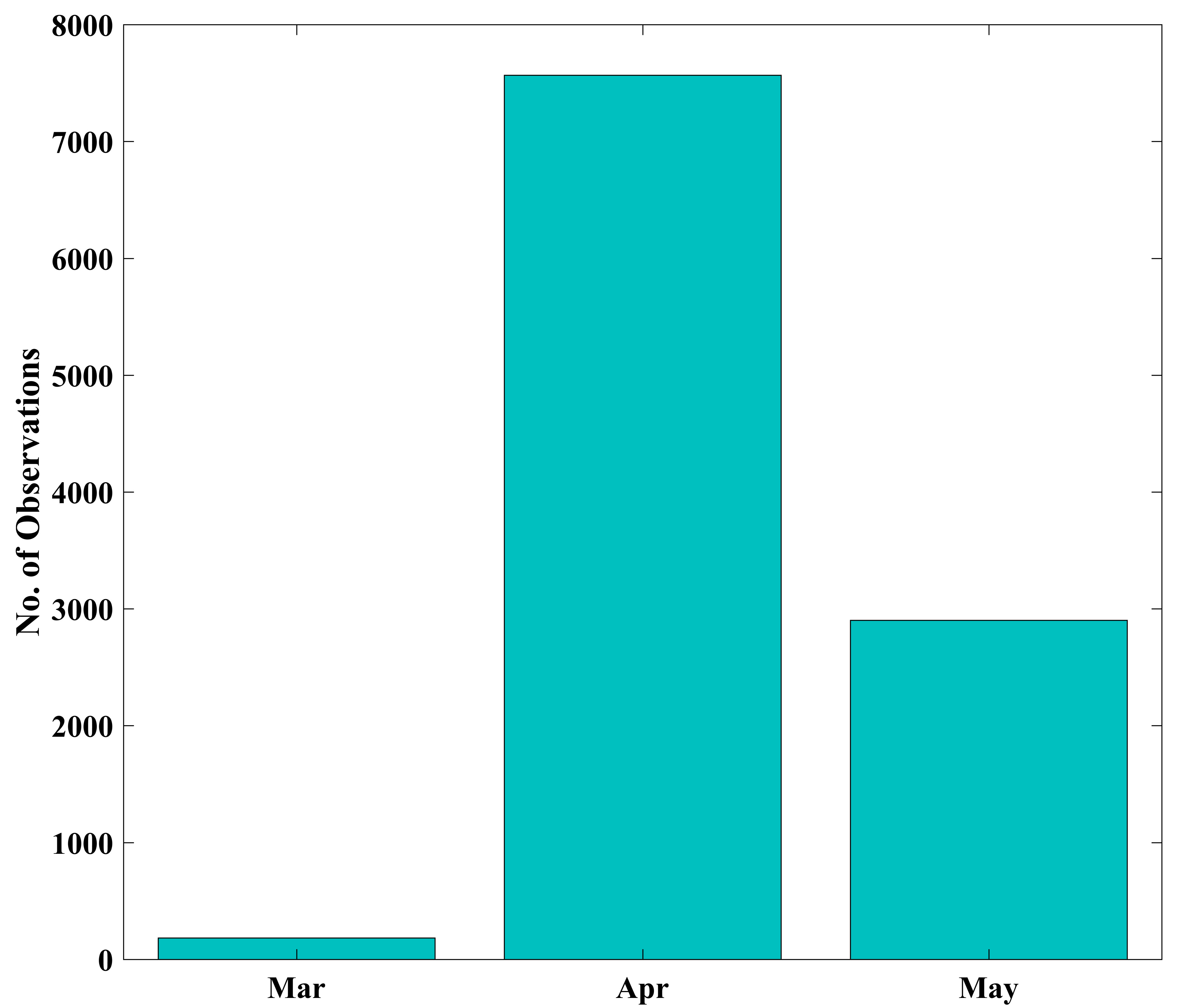}
\caption{Number of observations available for each month from the SOCAT cruise data.}
\label{fig:3}
\end{figure}

The RAMA buoy data is available on a 3 hourly temporal resolution, but following the previous studies it is converted to a daily scale. The total number of data available from RAMA buoy is 1557. The data available for each month is shown in Fig.\ref{fig:2}. Almost equal data is available for each month, reducing bias towards any particular month. The highest data is available in the month of May, followed by March and April. February has the lowest number of observations. The number of observations available from the SOCAT data (after removing values inside the 400 km from the coast) are 10652. The available monthly observations for SOCAT is shown in Fig.\ref{fig:3}. We observe that the highest observations are available for April. Moreover, only observations for three months is available, which suggests that the seasonality of the predictive machine learning model would be possible driven by the RAMA buoy data. SOCAT has limited spatial extent but enough to cover the present study domain (refer Fig.\ref{fig:1}).

\begin{figure}[ht!]
\centering
\includegraphics[height=8cm,width=\textwidth]{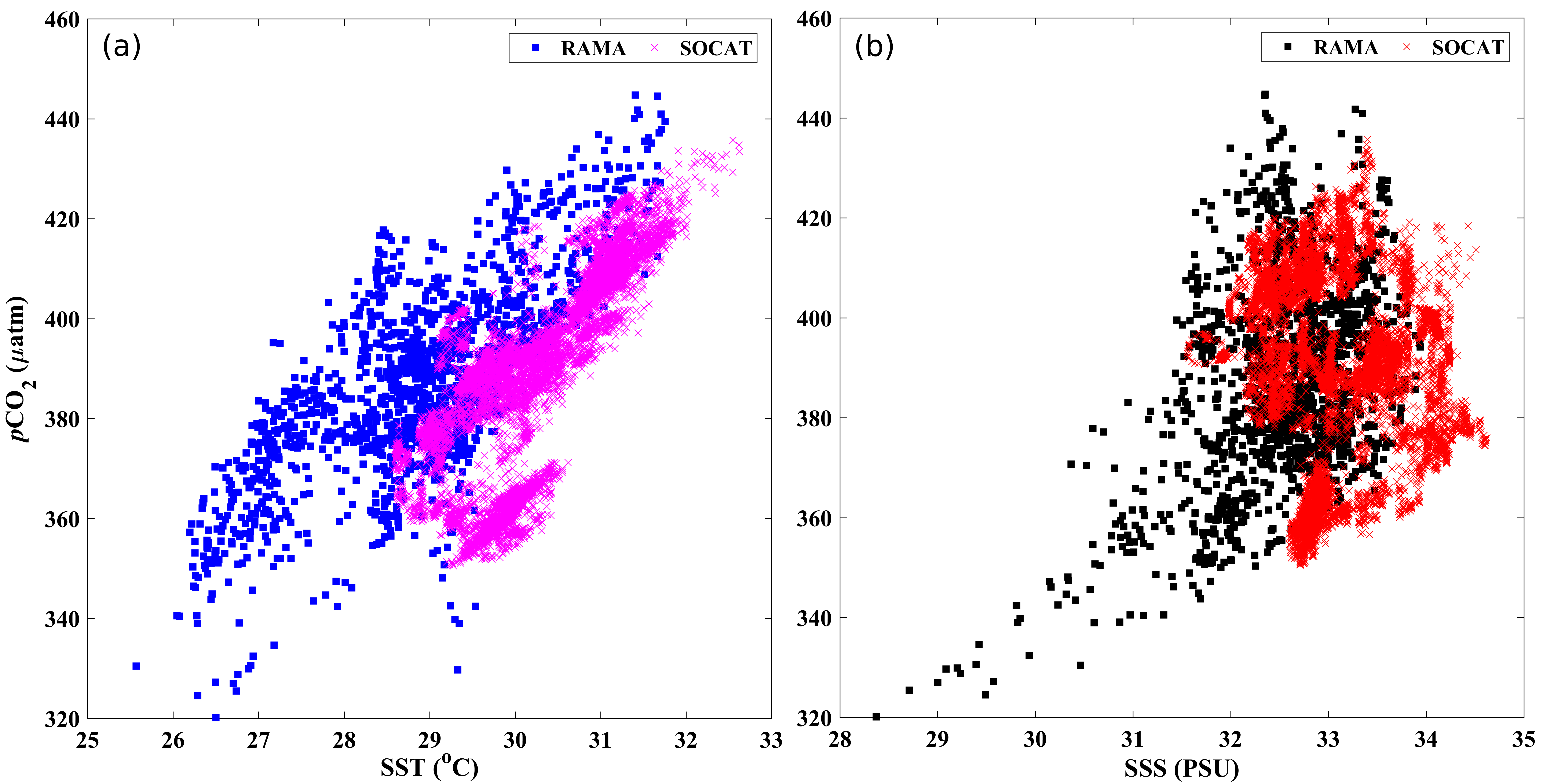}
\caption{The relationship of \textit{p}CO\textsubscript2 with the (a) SST and (b) SSS.}
\label{fig:4}
\end{figure}

In Fig.\ref{fig:4}, we analyze the relationship between the \textit{p}CO\textsubscript2 with the SST and SSS, respectively. The panel (a) of Fig.\ref{fig:4} demonstrates the relationship between the SST and the \textit{p}CO\textsubscript2. The blue dots represents the RAMA buoy data, and the pink crosses show the SOCAT data. A strong linear relationship is observed (Fig.\ref{fig:4}a) between the SST and the \textit{p}CO\textsubscript2 (r\textsuperscript{2} = 0.74 and r\textsuperscript{2} = 0.79 for RAMA and SOCAT, respectively, with p $<$ 0.001. However, the relationship between the SSS and the \textit{p}CO\textsubscript2 is linear but weaker for the RAMA buoy data (r\textsuperscript{2} = 0.42), and the linearity relinquishes while comparing the SOCAT derived SSS and \textit{p}CO\textsubscript2 (r\textsuperscript{2} = -0.0904, p $<$ 0.001). The possible explanation for the linear relation between the SSS and the \textit{p}CO\textsubscript2 for the RAMA derived data could be the lower salinity. The fresher waters contain lower \textit{p}CO\textsubscript2 but in the higher range of SSS values ($\geq$ 32 PSU) the linear relationship seizes to exist. 

\begin{figure}[ht!]
\centering
\includegraphics[height=8cm,width=\textwidth]{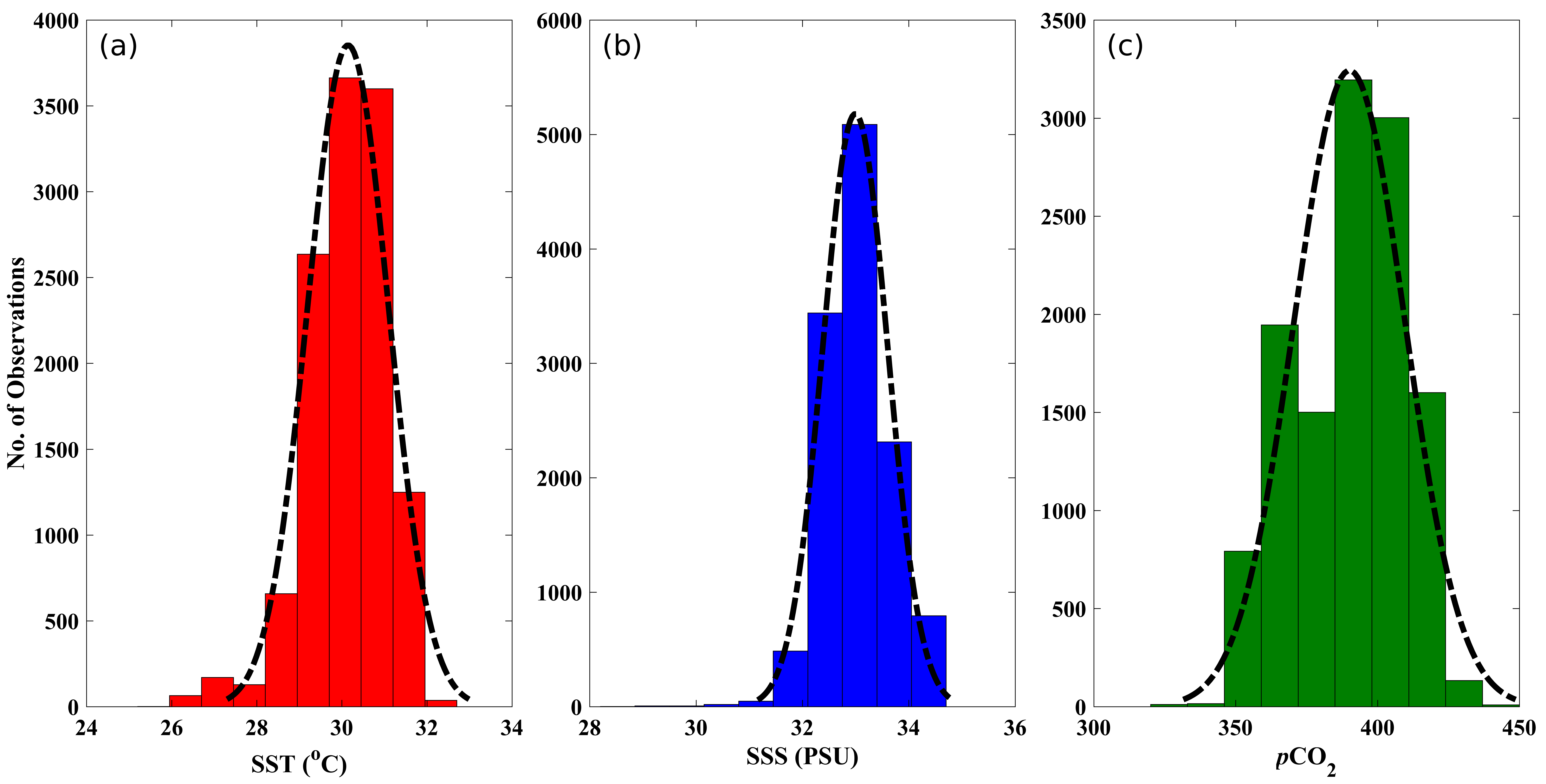}
\caption{Distribution of (a) SST, (b) SSS, and (c) \textit{p}CO\textsubscript2.}
\label{fig:5}
\end{figure}

Before splitting the data into training and testing, we visualize the distribution of the total collected raw data (total number of observations is 12208). Fig.\ref{fig:5} shows the distribution of all the variables (SST, SSS, and \textit{p}CO\textsubscript2) used in the present study. The distribution of all the variables in Fig.\ref{fig:5} indicates to be a normal distribution. The mean ($\mu$) and standard deviation ($\sigma$) for SST is 30.146 \textdegree{}C and 0.95 \textdegree{}C respectively, for SSS is 33.9 PSU and 0.61 PSU respectively, and for \textit{p}CO\textsubscript2 is 390.14 $\mu$atm and 19.52 $\mu$atm, respectively. The outliers ($>$ 3$\sigma$) are removed from the data, as many machine learning algorithms could be sensitive to these outliers. 

\subsubsection{Data splitting and scaling}
\label{s:2.2.2}

We split the data randomly into the training and testing set (80:20) using the train-test split from the sklearn library \citep{pedregosa2011scikit}. The early splitting aids in avoiding data leakage. The test data is only used for validation purposes. We use the same set of training and testing data for all the models, which enables us to compare the performance of these models with each other. For each of the models used in this study, while training the model, the training data is divided into 10 K-folds, and each of these folds are treated as the validation set. This K-fold technique helps in avoiding over-fitting.

After splitting the data, we scale the data using the StandardScaler method from the scikit-learn pre-processing library to ease the learning process for the models. The StandardScaler scales the distribution of data such that the mean of these data is 0 and the standard deviation is 1. This is also known as the standardization method, which uses the following formula:

\begin{equation}
    Z = \frac{(X - \mu)}{\sigma}
\end{equation}

\subsubsection{Models}
\label{s:2.2.3}

We implement three models or machine learning algorithms in the present study. Our goal is to first understand which model performs the best. The best model is then further fed with the satellite-derived data to examine the trend of sea-surface \textit{p}CO\textsubscript2 in the central BoB region. The description of the models used in this study are as follows:

\begin{itemize}
    \item \textbf{Multiple Linear Regression (MLR)} :\\
    Multiple Linear regression (MLR) is an inferential technique that determines the dependent variables based on the independent variables. The technique is aimed to establish linear relationship between the interpretive and the response variables. It is an extension of the ordinary least square method as it involves more than one revelatory variable. The mathematical formulation for the MLR is :
    \begin{equation}
        Q_i  = a_o + {a_1}{P_{i1}} + {a_2}{P_{i2}} + ... + {a_h}{P_{ih}}
    \end{equation}
    where for i = n observations:\\
    $Q_i$ = dependent variable\\
    $P_i$ = independent variables\\
    $a_o$ = y-intercept (constant)\\
    $a_h$ = slope coefficient for each independent variable\\
    The MLR technique has been a preferred model to study the sea-surface \textit{p}CO\textsubscript2 in the BoB region \citep{dixit2019net,sridevi2021role,mohanty2022surface}. Observing the data in Fig.\ref{fig:4}, a weaker linear relationship between SSS and \textit{p}CO\textsubscript2 may weaken the performance of this model. Hence we are encouraged to examine the performance of some other popular models hypothesising better predictability of the sea-surface \textit{p}CO\textsubscript2. To execute the MLR, we employ the linear regression class from the scikit-learn library. The regression fit for predicting the sea-surface \textit{p}CO\textsubscript2 is:
    
    \begin{equation}
        \textit{p}\text{CO\textsubscript2} = 390.71 + 13.37\text{SST} + {0.92}\text{SSS}
    \end{equation}
    
    \item \textbf{Artificial Neural Network (ANN)} :\\
    \begin{figure}[ht!]
    \centering
    \includegraphics[height=10cm,width=\textwidth]{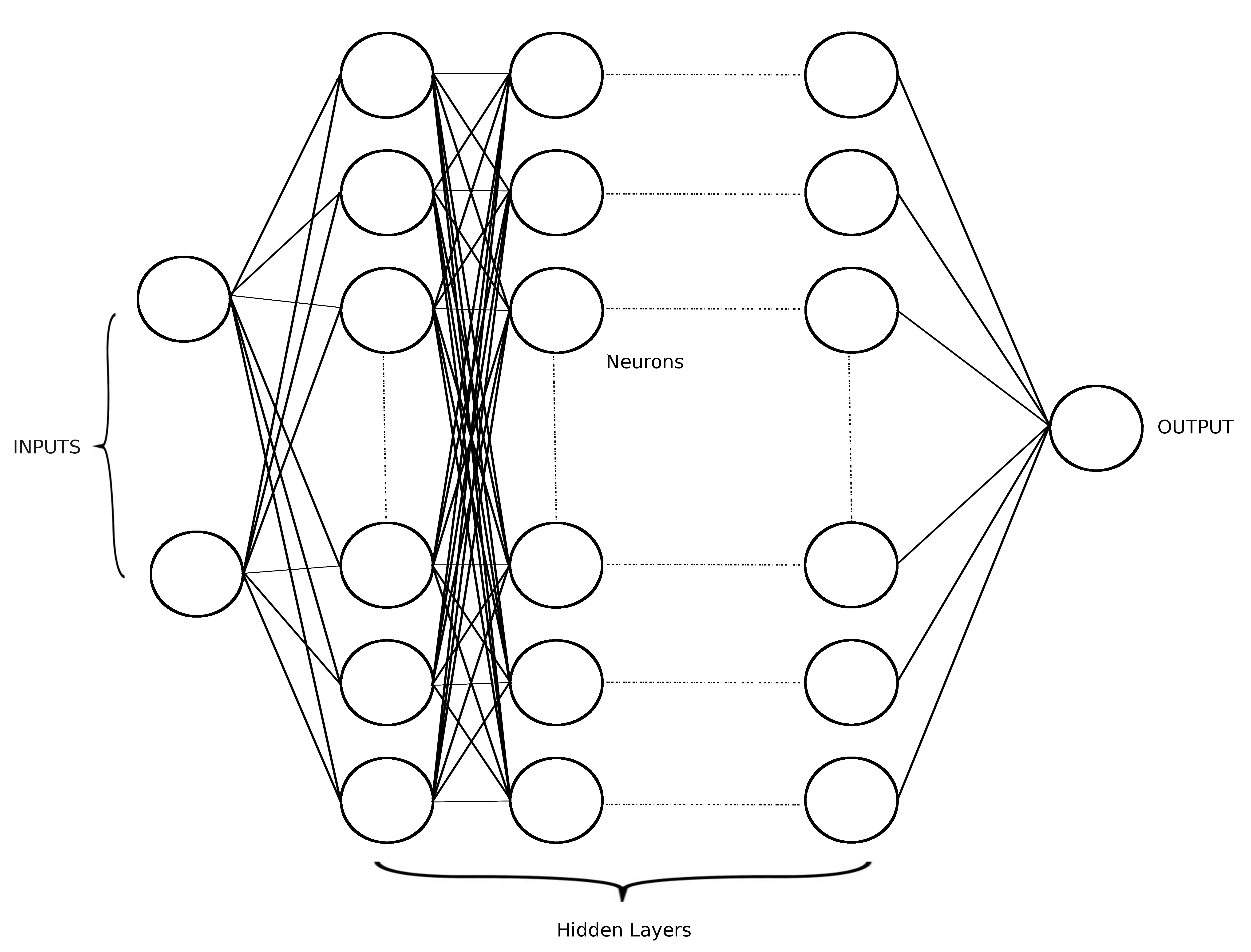}
    \caption{Schematic representation of the ANN model.}
    \label{fig:6}
    \end{figure}
    
    The Artificial Neural Network (ANN) is a mathematical computational model which is inspired by the network in the animal brain. Like the brain, the ANN consists of neurons, which are responsible for processing the incoming signals. The \say{signal} at each connections are real numbers, and the output of each neuron is computed as a sum of the inputs to neurons which is a non-linear function. The connection between the neurons are also known as the edges. The neurons and edges are assigned weights, which adjusts itself to an optimum output. A signal travelling from the input layer to the output layer may traverse through a bunch of hidden layers, each consisting of several neurons. Fig.\ref{fig:6} is a depiction of the ANN applied in this study. The two inputs in Fig.\ref{fig:6} shows the input parameters (SST and SSS), and the network's output is the sea-surface \textit{p}CO\textsubscript2.
    The number of hidden layers, neurons in each layer, and the learning rate are the tunable hyper-parameters chosen for this study.\\
    
    \begin{table}[ht!]
        \centering
        \begin{tabular}{|c|c|}
        \hline
        \textbf{Hyper-parameters}  & \textbf{Range/Choices}  \\
        \hline
        \hline
        Hidden Layers & 2-20 \\
        \hline
        Neurons per layer & 20-60 (step=3) \\
        \hline
        Learning Rate & 0.01, 0.001, 0.0001 \\
        \hline
        \end{tabular}
        \caption{Search space for each of the hyper-parameters.}
        \label{tab:2}
    \end{table}
    
    We use the KerasTuner \citep{o2019keras} class from the Keras library to find the optimum values of the ANN hyper-parameters. Table \ref{tab:2} explicitly demonstrates the range for each of the hyper-parameters. We use the RandomSearch class from the KerasTuner library to tune the hyper-parameters. We use the Rectified Linear Unit (ReLU) \citep{agarap2018deep} activation function in the hidden layers and \say{Linear} activation function in the output layer. The \say{\text{he-uniform}} \citep{he2015delving} technique is used to initialize the weights. We choose the \say{Mean Absolute Error} as the loss function, which has to be minimized. The parameters are tuned for 100 maximum trials, with 2 execution per trial.
    
    \begin{table}[ht!]
        \centering
        \begin{tabular}{|c|c|}
        \hline
        \textbf{Hidden Layer} & \textbf{Number of neurons}  \\
        \hline
        \hline
        Layer 1 & 32 \\  
        \hline
        Layer 2 & 29 \\  
        \hline
        Layer 3 & 56 \\  
        \hline
        Layer 4 & 53 \\  
        \hline
        Layer 5 & 47 \\  
        \hline
        Layer 6 &  44\\  
        \hline
        Layer 7 &  56\\  
        \hline
        Layer 8 &  20\\  
        \hline
        Layer 9 &  32\\  
        \hline
        Layer 10 &  23\\  
        \hline
        Layer 11 &  20\\  
        \hline
        Layer 12 &  50\\  
        \hline
        Layer 13 &  47\\  
        \hline
        Layer 14 &  23\\  
        \hline
        Layer 15 &  47\\  
        \hline
        Layer 16 &  23\\  
        \hline
        Layer 17 &  32\\  
        \hline
        \end{tabular}
        \caption{Number of neurons in each hidden layer.}
        \label{tab:3}
    \end{table}
    
    After the hyper-parameter optimization, we attain the final values for the ANN. The ANN for this study has a total of 17 hidden layers. The number of neurons corresponding to each of the hidden layer is shown in Table \ref{tab:3}. The model performs best at a learning rate of 0.001. With this final structure of the model, we run the ANN for 1000 epochs.
    
    \item \textbf{Xtreme Gradient Boosting (XGBoost)} :\\
    
    \begin{figure}[ht!]
    \centering
    \includegraphics[height=10cm,width=\textwidth]{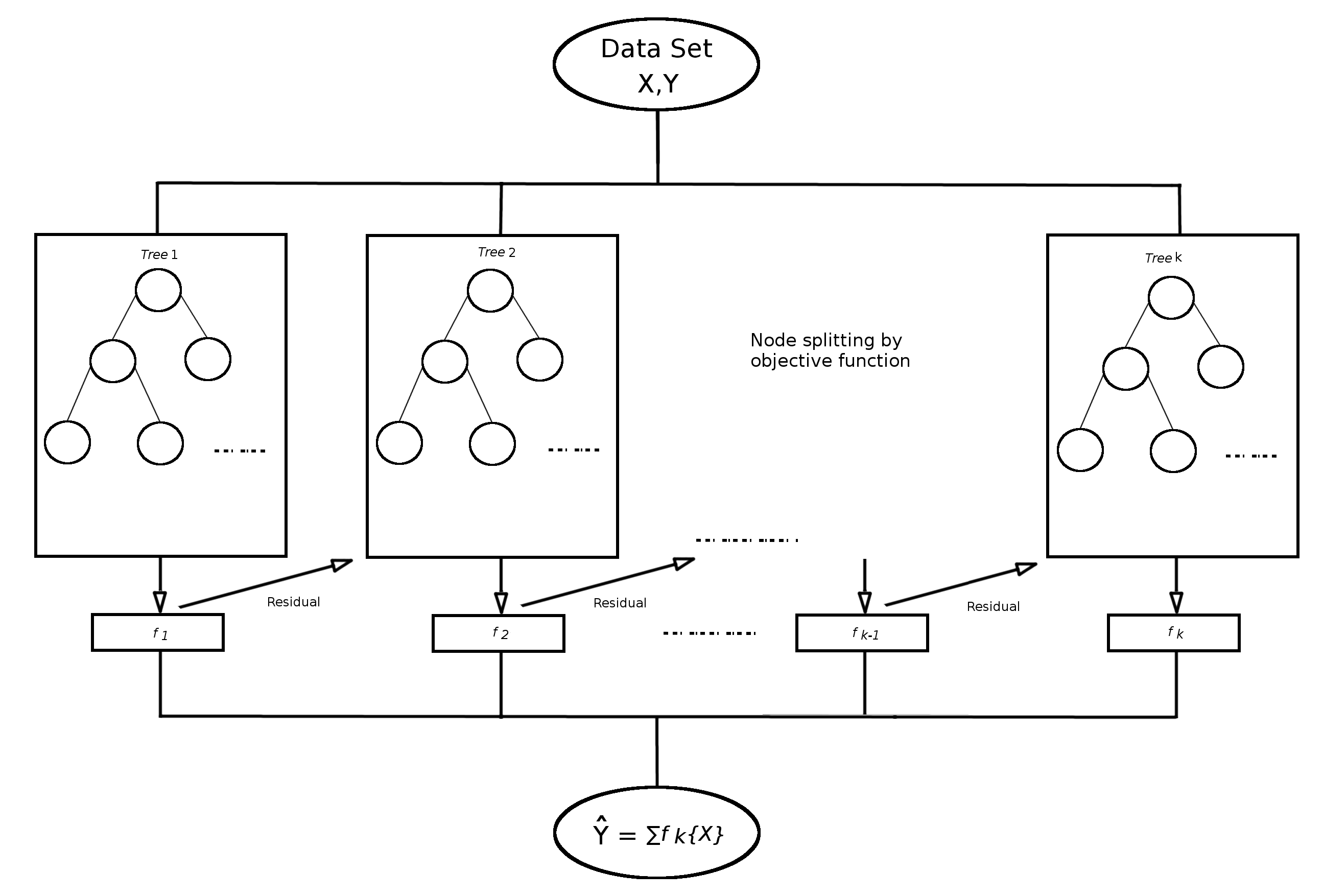}
    \caption{Schematic representation of the XGBoost model.}
    \label{fig:7}
    \end{figure}

    Xtreme Gradient Boosting (XGBoost) \citep{chen2016xgboost} belongs to the family of the decision tree based boosting algorithms. The only purpose for developing the XGBoost algorithm is to expand the performance along with the computational speed for the gradient boosted algorithm. The XGBoost model is trendy among the Kagglers \url{https://www.kdnuggets.com/2017/10/xgboost-top-machine-learning-method-kaggle-explained.html}. The high speed and high accuracy of the model has encouraged us to examine its performance alongside the ANN. Fig.\ref{fig:7} emulates a schematic representation of the XGBoost algorithm. The trees or weak learners are basically added in sequential order, only the residuals are fed to the next weaker learner. This approach aids in reducing the errors. The Newton boosting based on the Newton Raphson method, unlike gradient descent, approaches global minima in an accelerated manner. The generic algorithm for XGBoost is as follows:\\
    
    Let us consider a training set having $n$ number of observations,\\
    \begin{equation}
        M = (x_1,y_1),(x_2,y_2),(x_3,y_3),..,(x_n,y_n)
    \end{equation}
    
    let us consider a loss function $J(y,F(x))$ and assume there is $K$ number of weak learners. First, we initialize the model with a constant value:
    \begin{equation}
        \hat{G}_{(0)}(x) = \text{argmin}{\sum_{1}^{n}}{J(y_i,t)}
    \end{equation}
    
    Further, for each of the sequentially added models or weak learners ($k$ = 1 to $K$), the loss function needs to be minimized. To reduce the loss function, Taylor approximation is implemented, which requires the first ($\hat{s_k}(x_i)$) and second ($\hat{v_k}(x_i)$) order derivative of the loss function. A base learner is then fitted by solving the optimization problem (Eq.\ref{e:6}), which is then multiplied with the learning rate ($\zeta$) to find the final solution for a particular base learner (Eq.\ref{e:7}). The final model is then updated by adding the just found optimized solution (Eq.\ref{e:8}). The final output of the XGBoost is the ensemble of all the outputs from each of the weak learners (Eq.\ref{e:9}).
    
    \begin{equation}
        \hat{\Theta_k} = \text{argmin}{\sum_{1}^{n}}\frac{1}{2}{\hat{v_k}(x_i)}\left[-\frac{\hat{s_k}(x_i)}{\hat{v_k}(x_i)} - \Theta(x_i)\right]^2
    \label{e:6}
    \end{equation}
    
    \begin{equation}
        \hat{G}_{k}(x) = \zeta \times \hat{\Theta}_{k}(x)
    \label{e:7}
    \end{equation}
    
    \begin{equation}
        \hat{G}_{k}(x) = \hat{G}_{(k-1)}(x) + \hat{G}_{k}(x)
    \label{e:8}
    \end{equation}
    
    \begin{equation}
        \hat{G}(x) = \sum_{k=1}^{K}\hat{G}_{k}(x) 
    \label{e:9}
    \end{equation}
    
    Similar to the ANN, the XGBoost model, also has tunable hyper-parameters. In this study, we employ the Optuna optimization framework \citep{akiba2019optuna}. The description of all the XGBoost parameters is available at: \url{https://xgboost.readthedocs.io/en/stable/parameter.html}, hence it is not repeated here. The range and the final optimized values of the hyper-parameter are shown in Table \ref{tab:4}.
 
    \begin{table}[ht!]
        \centering
        \begin{tabular}{|c|c|c|}
        \hline
        \textbf{Hyper-parameters}  & \textbf{Range} & \textbf{Optimized Value} \\
        \hline
        \hline
        lambda & 0.8-1.0 & 0.99\\
        \hline
        alpha & 0-1 & 9.99e-21\\
        \hline
        subsample & 0.2-1 & 0.99\\
        \hline
        colsample\_bytree & 0.2-1 & 0.99\\
        \hline
        max\_depth & 1-20 (step=1) & 6\\
        \hline
        min\_child\_weight & 1-50 & 1\\
        \hline
        learning\_rate & 0.001-0.3 & 0.3\\
        \hline
        gamma & 0-1 & 9.99e-08\\
        \hline
        n\_estimators & 50-500 (step=50) & 200\\
        \hline
        \end{tabular}
        \caption{Range and final values of the XGBoost hyper-parameters.}
        \label{tab:4}
    \end{table}   
\end{itemize}
 
\section{Results and discussions}
\label{s:3}

\subsection{Model Selection}
\label{s:3.1}

In the previous section, we discuss the data acquisition and explore the characteristics of the data through data visualization. Further, after randomly splitting the data, we run three models (MLR, ANN, and XGBoost). In this section, we evaluate the performance of each of these models against the same test data. To facilitate the evaluation process, we use the Taylor diagram visualization \citep{taylor2001summarizing}. The Taylor diagram enables summarised visualisation of the model performance using three statistical indices. These statistical indices are: Correlation coefficient (r), Standard Deviation (STD), and Root Mean Square Error (RMSE). The correlation coefficient ranges between -1 to 1 and highlight the linear behaviour between the model output and the observations. The standard deviation measures the deviation of data from its mean. Ideally the STD of the model output should be equal to the STD of the observation. The error in the model predictability is known through RMSE. A lower RMSE indicates better model prediction.

\begin{figure}[ht!]
\centering
\includegraphics[height=10cm,width=10 cm]{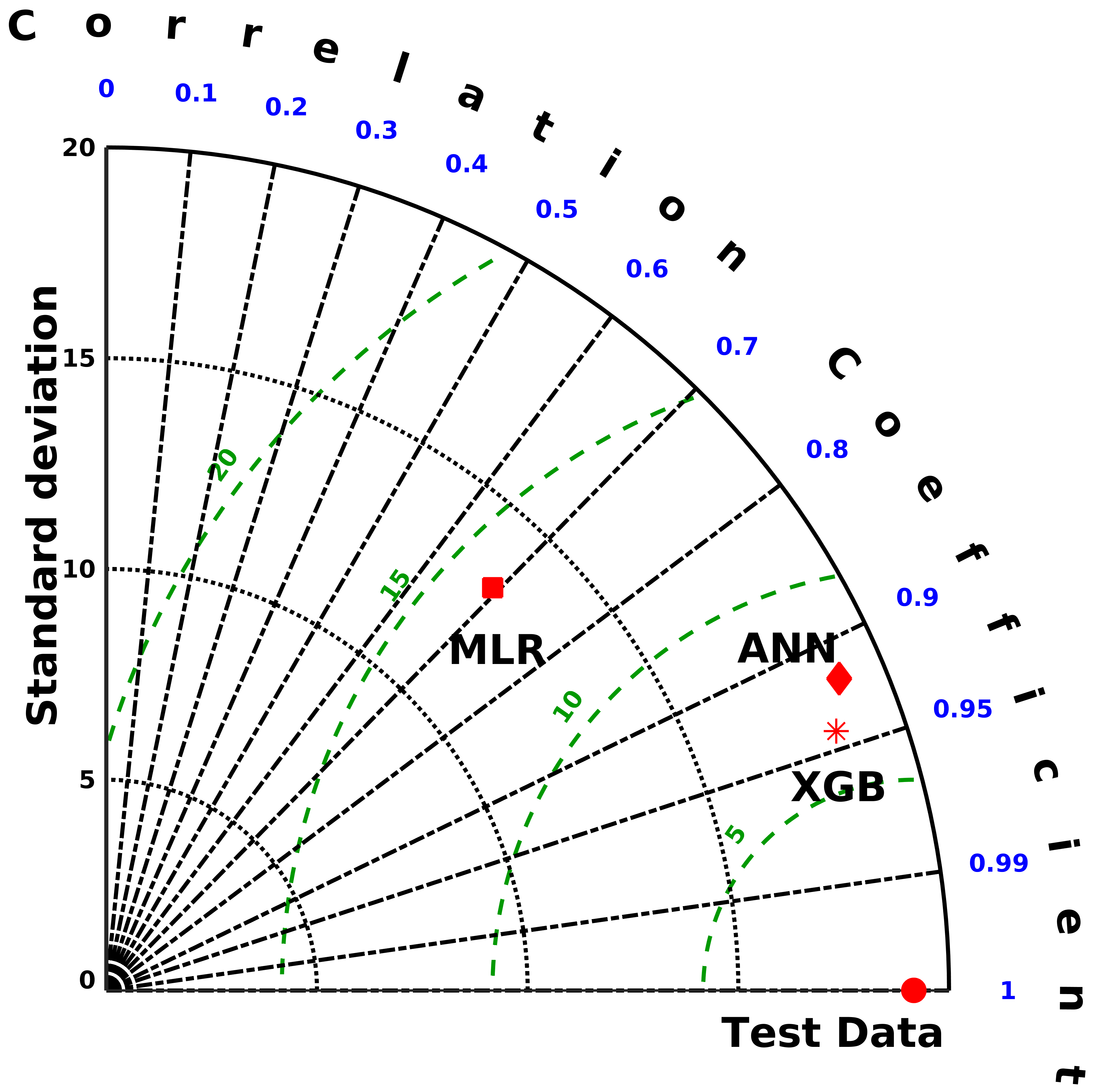}
\caption{Taylor diagram represents the model performance against the test data.}
\label{fig:8}
\end{figure}

Fig.\ref{fig:8} shows the performance of each of the models used in the present study. The X-axis and Y-axis shows the STD of the models, whereas the correlation coefficient is along the angular arc. The green dashed lines indicate RMSE. The test data lies on the X-axis as it is completely correlated with itself (r = 1) and has zero errors. The red square in Fig.\ref{fig:8} represents the performance of the MLR model. The MLR model has a correlation of 0.69, which indicate that the model output is poor in capturing the variation of the test data. The test data has a STD of 19.16 $\mu$atm, but the MLR model output has a STD of 13.24 $\mu$atm, which is a drift apart from the test data. The RMSE of the MLR model output is $\pm$13.83 $\mu$atm, which is relatively large.

The performance of the ANN and the XGBoost (red diamond and red asterisk, respectively in Fig.\ref{fig:8}) is far better than the MLR model. The ANN shows an excellent correlation (r = 0.92) with the test data. The STD and RMSE of the ANN model is 18.90 $\mu$atm and $\pm$7.61 $\mu$atm, respectively. This indicates a good performance by the ANN model. The XGBoost model has a better correlation (r = 0.94) with the test data than any other two models. The RMSE of the XGBoost algorithm ($\pm$6.42 $\mu$atm) is the lowest among all the models used in this study. The STD of the XGBoost model (18.4 $\mu$atm) is less than that of ANN, but the difference is negligible. Hence, Fig.\ref{fig:8} noticeably shows that the XGBoost model is the best performing model, and the MLR model has the worst performance. The lower linearity between the SSS and the sea-surface \textit{p}CO\textsubscript2 could be a potential explanation of the poor performance of the MLR model. The marginally better performance of the XGBoost algorithm may be attributed to less number of data and possibly features. In the study of sea-surface \textit{p}CO\textsubscript2, the luxury of collecting large amount of data is missing (especially in the BoB), hence XGBoost may be considered as a better model even with more features. 

\subsection{Predicting sea-surface \textit{p}CO\textsubscript2 using satellite data}
\label{s:3.2}

\begin{figure}[ht!]
\centering
\includegraphics[height=8cm,width=\textwidth]{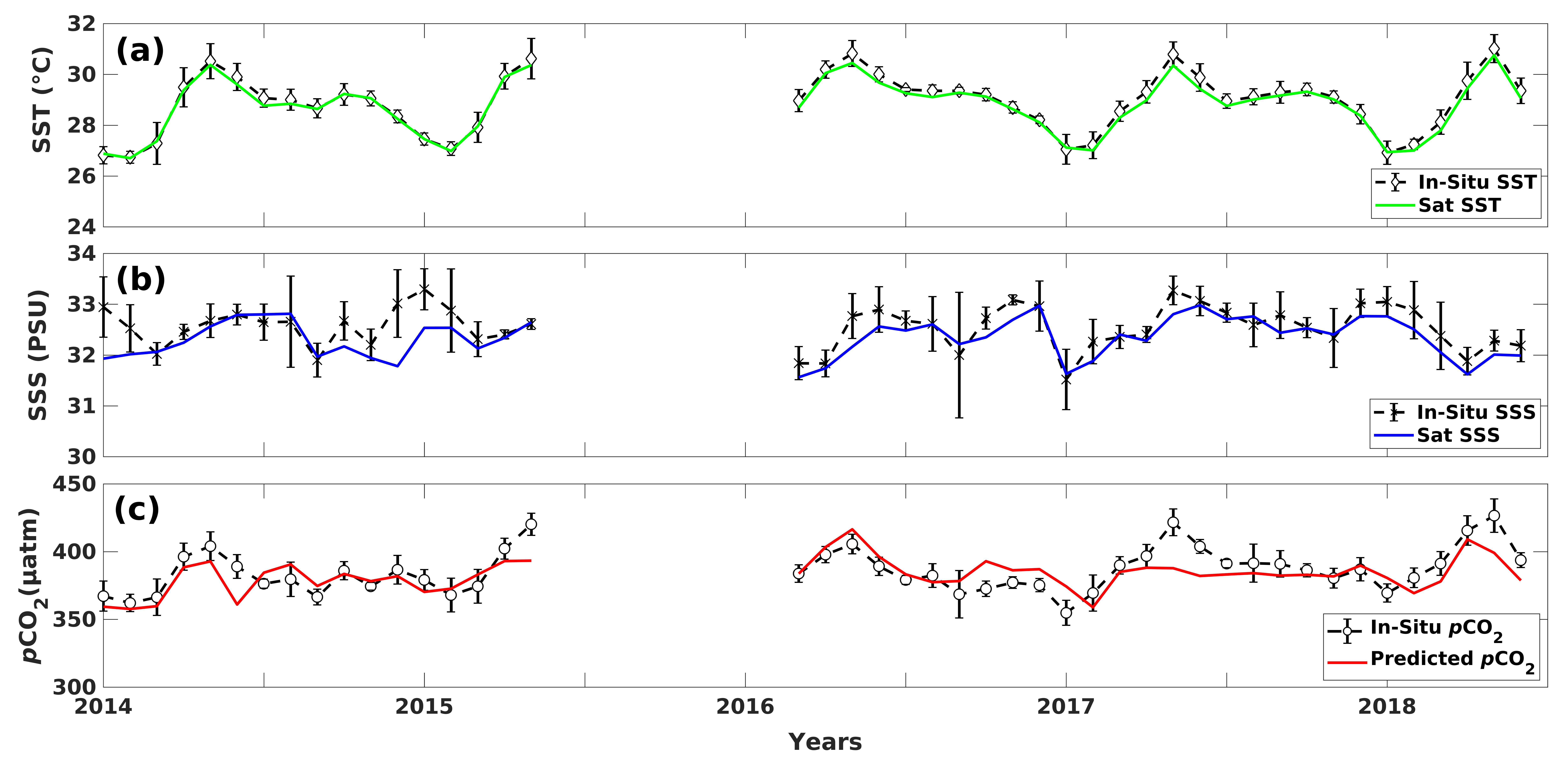}
\caption{Comparison of the interannual variation of the satellite-derived (a) SST and (b) SSS with the in-situ observations from the RAMA buoy. In Fig.(c) the predicted sea-surface \textit{p}CO\textsubscript2 using XGBoost model is compared with the in-situ RAMA buoy data on an interannual scale. }
\label{fig:9}
\end{figure}

The XGBoost model is the best performing model is well established in the previous section. In this section, we would look to analyze the potentiality of the XGBoost model to mimic the sea-surface \textit{p}CO\textsubscript2 using the satellite-derived SST and SSS. The source of the SST and SSS is already mentioned in Sec.\ref{s:2.1.3}. Before scrutinizing the performance of the XGBoost model, it is important to analyze the satellite-derived SST and SSS. 

\begin{figure}[ht!]
\centering
\includegraphics[height=13cm,width=\textwidth]{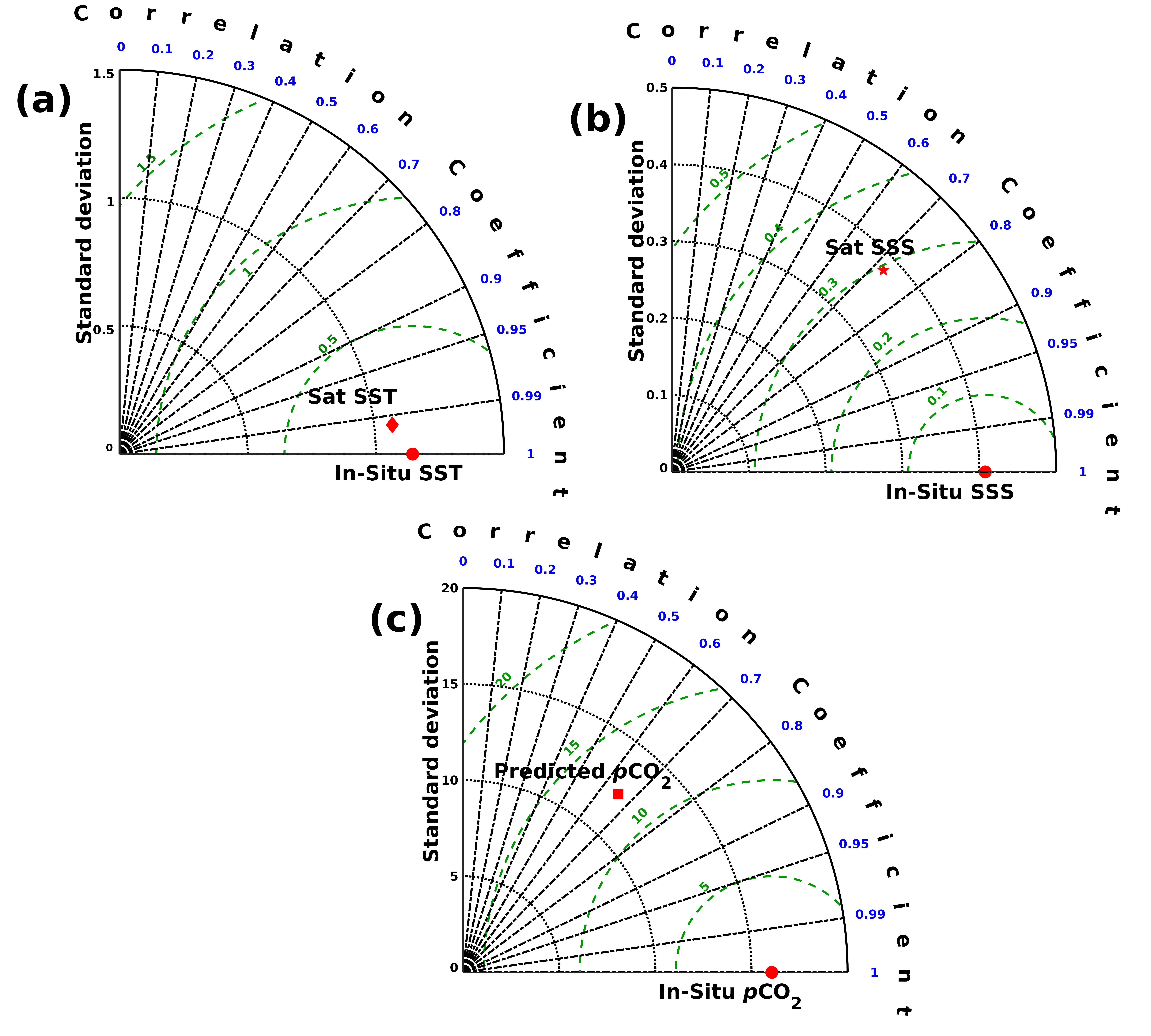}
\caption{Taylor diagram represents the performance of the satellite (a) SST, (b) SSS, and (c) the predicted sea-surface \textit{p}CO\textsubscript2 with the in-situ RAMA buoy data.}
\label{fig:10}
\end{figure}

In Fig.\ref{fig:9}a and b, we compare the SST and SSS from the satellite with the in-situ observations. The SST seems to have a close match with the observations (Fig.\ref{fig:9}a), the crests and troughs of the observed SST is excellently emulated by the satellite data (r = 0.99). The bias between both the data are almost negligible. The Taylor diagram in Fig.\ref{fig:10}a advocates for the excellent performance of the satellite SST with the in-situ observation. The STD of the in-situ SST is 1.14 \textdegree{}C, and the satellite-derived SST is 1.07 \textdegree{}C. It has an overall RMSE of $\pm$0.14 \textdegree{}C, which suggests that the satellite-derived SST contains trivial errors. 

The comparison between the satellite-derived SSS and the in-situ SSS from RAMA buoy, reveals a good agreement between the two. The satellite-derived SSS utilized in this study is the one of the best SSS product for the BoB region \citep{akhil2020bay}, and yet the Fig.\ref{fig:9}b shows that the satellite-derived SSS still has some potential to improve. Fig.\ref{fig:9}b indicates that the satellite SSS has a significant bias in the second half of 2014 and January-March 2015. Though from March 2016 to June 2018, the satellite SSS reasonably emulates the is-situ SSS, though it seems to have a minor underestimation. Fig.\ref{fig:10}b represents the statistical performance of the satellite SSS product. The correlation of 0.73 suggests the satellite SSS has captured the interannual monthly variation in a satisfactory manner. The STD of the satellite (0.38 PSU) and the in-situ SSS (0.41 PSU) lies in close proximity. The RMSE of $\pm$0.29 PSU between the satellite SSS and observations divulge a small but acceptable error between the two datasets. 

Fig.\ref{fig:9}c compares the modeled sea-surface \textit{p}CO\textsubscript2 with the in-situ sea-surface \textit{p}CO\textsubscript2. The XGBoost model seems to have performed reasonably in emulating the \textit{p}CO\textsubscript2 of the BoB region. The model captures the interannual monthly variation satisfactorily except for 2017. The peak in the April-June of 2017 is not captured by the model, also a lag in May 2018 is a notable difference between the model and in-situ data. The errors in the SSS could be a possible reason for the deviation of the modeled sea-surface \textit{p}CO\textsubscript2 from the observations. The statistical inference of the model performance is displayed in Fig.\ref{fig:10}c. The modeled sea-surface \textit{p}CO\textsubscript2 displays an average correlation of 0.65 when compared with the in-situ data. The low correlation between the satellite SSS and the observation could be one of the reasons for the average correlation of the modeled \textit{p}CO\textsubscript2. The STD of the in-situ \textit{p}CO\textsubscript2 data is 16.05 $\mu$atm, whereas the STD in the modeled \textit{p}CO\textsubscript2 is 12.5 $\mu$atm, the low difference between the STDs of each of these data shows that the model closely emulates the variance. The RMSE of $\pm$12.23 $\mu$atm is reasonably low.

Since we compare the modeled sea-surface \textit{p}CO\textsubscript2 in this study with the available RAMA buoy data, it seems interesting to compare the results with the available results of some recent used model studies \citep{sridevi2021role,mohanty2022surface}. The modeled \textit{p}CO\textsubscript2 from the \cite{sridevi2021role} is evaluated with the RAMA buoy for the period of November 2013 to June 2015, hence we choose the period of January 2014 to June 2015, which are two months less (practically we are avoiding just two values) and then compare the performance of the models of the two studies. \cite{sridevi2021role} has mentioned that the RMSE between the MLR model and in-situ data for their study to be $\pm$16 $\mu$atm, whereas with the implementation of the XGBoost model we are able to reduce this RMSE to $\pm$11.01 $\mu$atm. Similarly, the period of model performance analysis for the \cite{mohanty2022surface} is December 2013 to November 2014, so to keep the period of analysis as similar as possible, we choose the time frame of January-December 2014 (i.e., the whole year 2014). The correlation coefficient (r = 0.75) of the MLR model of \cite{mohanty2022surface} is fairly close to that revealed by the XGBoost model (r = 0.7) used in this study. The RMSE from the study of \cite{sridevi2021role} is $\pm$38 $\mu$atm, and the RMSE from this study for a similar time period is $\pm$10.2 $\mu$atm. This shows that the XGBoost model has the capability of reproducing the sea-surface \textit{p}CO\textsubscript2 better than some of the recent studies, also it expands the applicability of the XGBoost model in predicting the sea-surface \textit{p}CO\textsubscript2 of the BoB region. 

\subsection{Trend analysis and seasonal variation in the SST, the SSS, and the sea-surface \textit{p}CO\textsubscript2 for the past decade}
\label{s:3.3}

\begin{figure}[ht!]
\centering
\includegraphics[height=7cm,width=\textwidth]{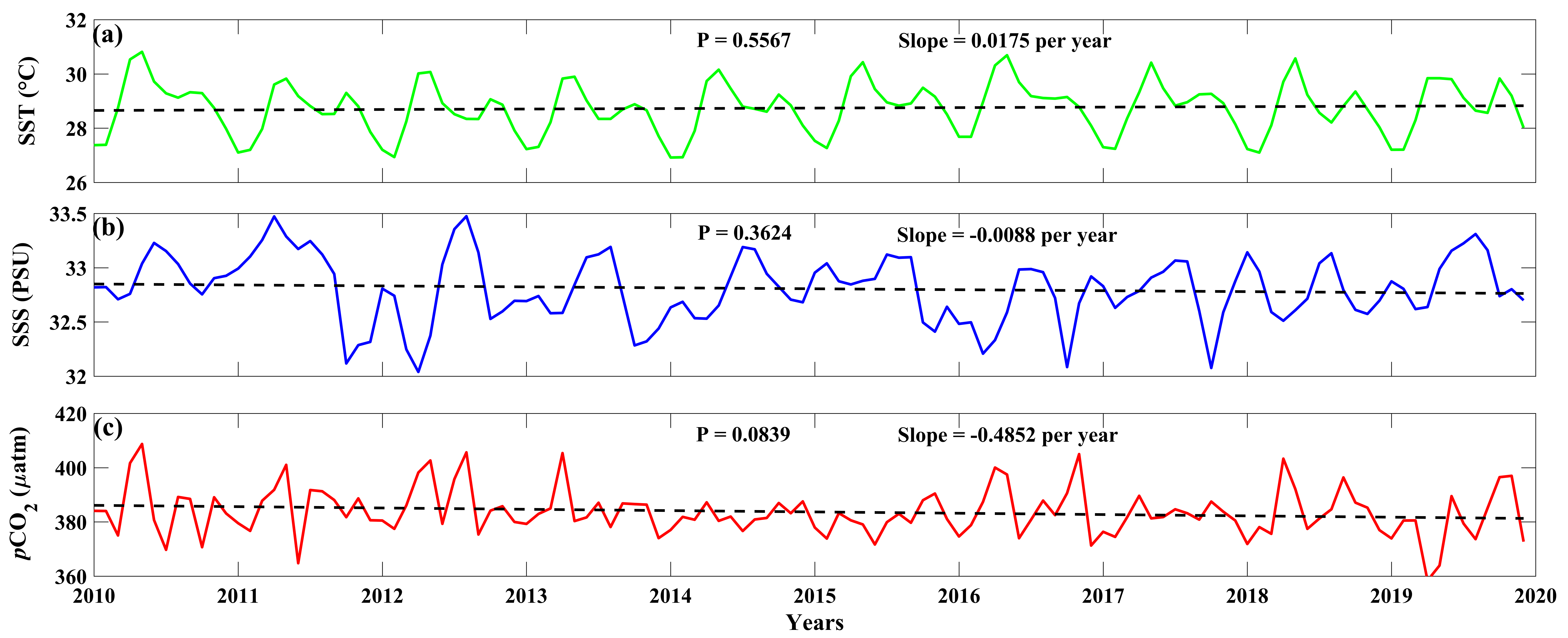}
\caption{Monthly mean variability in the (a) SST, (b) SSS, and (c) \textit{p}CO\textsubscript2 in the central BoB between 2010-2019. The P-value and the slope are depicted in each figure. }
\label{fig:11}
\end{figure}

After establishing the proficiency of the XGBoost model in the previous section, we now apply it with satellite-derived SST and SSS in the central BoB region (Fig.\ref{fig:1}). Our interest lies in examining and analyzing the trends of the SST, SSS, and the sea-surface \textit{p}CO\textsubscript2 in central BoB in between 2010-2019 (last decade). Fig.\ref{fig:11} exhibits the inter-annual monthly variation (domain mean) of SST, SSS, and sea-surface \textit{p}CO\textsubscript2. Fig.\ref{fig:11}a shows monthly variability of the SST in the last decade. The SST follows a repeated annual cycle throughout the last decade. An annual cycle has two high peaks \citep{colborn1975thermal}, one in May and the other in September. The maximum SST is around May each year, after which a steady decrease is observed till June. The SST has lower maxima around September, after which SST decreases until January. The trend reveals the warming of the ocean surface at a rate of 0.0175 \textdegree{}C per year (P = 0.56) in the last decade. This rise in SST could be attributed to global warming \citep{sridevi2021role}. The rate of increase of SST found in this study is inline with the studies of \cite{sridevi2021role}. The study of \cite{sridevi2021role} was from 1998 to 2015, a slight discrepancy (very low) in the rising rate of SST for the central BoB region could be due to the difference in the definition of the central BoB. 

The BoB receives a high amount of freshwater flux from rivers and the precipitation. This freshwater makes the BoB stratified, which inhibits vertical mixing and eventually the air-sea interactive processes. The freshwater spreads over the surface in the form of freshwater plume and reduces the sea-surface \textit{p}CO\textsubscript2 levels \citep{joshi2021influence,sridevi2021role}. The Ganga-Brahmaputra riverine system, which has one of the highest discharge rates in the world, originates from the Himalayas. Due to the rise in global warming (rise in global temperature by 0.18 \textdegree{}C per decade since 1981 \citep{lindsey2020climate}) the ice glaciers and the ice cover over the Himalayas has decreased significantly \citep{goes2020ecosystem}. This melting of the Himalayan glaciers result in a rise in the freshwater influx in the BoB region which consequently affects the stratification \citep{trott2019large}. Historically, the salinity in the BoB increased from 1940 to 1972, but from 1973 to 2008, the salinity is reported to be consistently decreasing. This reduction of salinity is attributed to the upsurge in precipitation and freshwater influx \citep{kumar2016comparison,mukherjee2018increase,goes2020ecosystem}.

The satellite SSS data shows a consistent decrease in the SSS through the past decade (Fig.\ref{fig:11}b). The SSS in the central BoB decreases at the rate of -0.0088 per year, with almost 70\% significance level. As discussed earlier, this slow but decline in SSS could be attributed to the melting of ice and the increased precipitation in the BoB. Unlike the SST, no consistent monthly variation pattern is observed for the SSS. The absence of any pattern in the SSS variability is probably responsible for a less recognizable cycle in the sea-surface \textit{p}CO\textsubscript2 (Fig.\ref{fig:11}c). The modeled sea-surface \textit{p}CO\textsubscript2 shows a decreasing trend in the central BoB region during the past decade. The \textit{p}CO\textsubscript2 decreases at a rate of -0.4852 $\mu$atm per year, approximately 90\% significance. 

\begin{figure}[ht!]
\centering
\includegraphics[height=8cm,width=\textwidth]{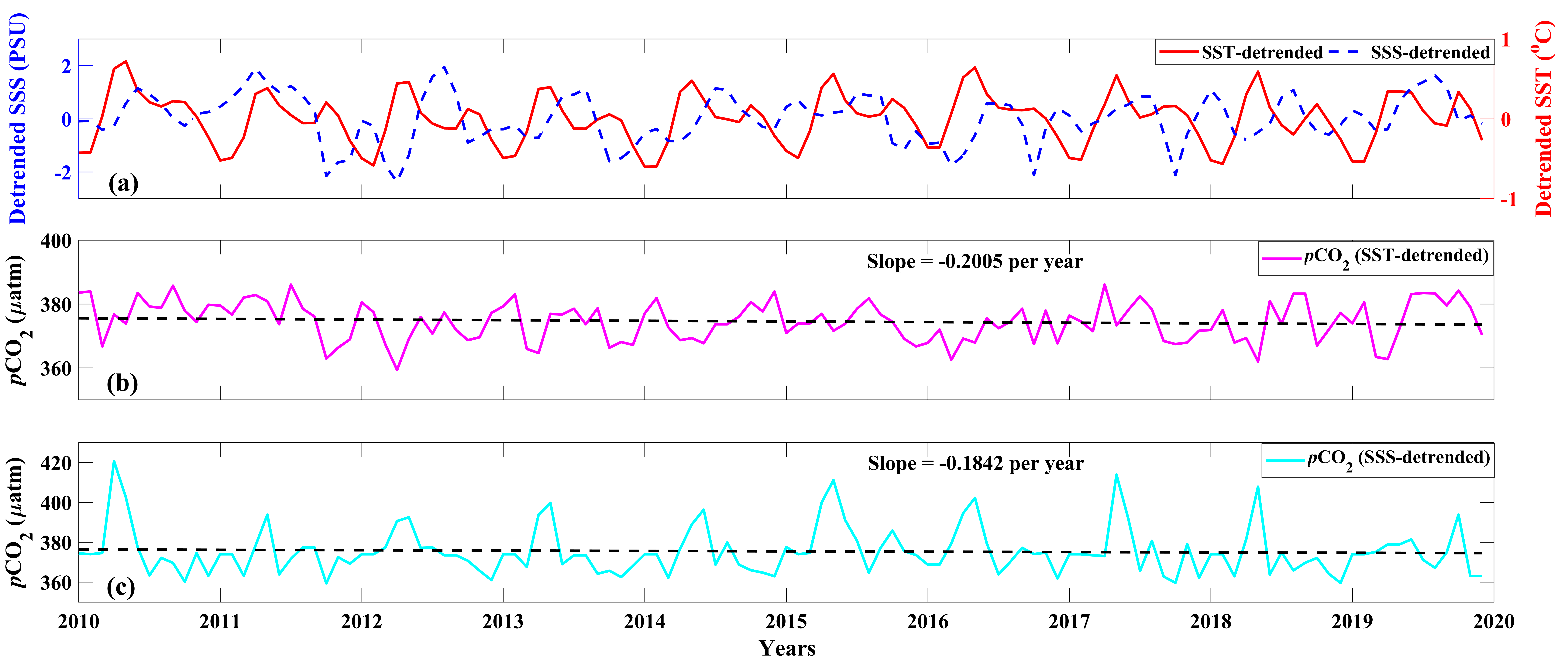}
\caption{The detrended SST and SSS is displayed with red continuous line and broken blue line in panel (a). The panel (b) and (c) shows the monthly variation in modeled \textit{p}CO\textsubscript2 using the detrended SST and SSS, respectively.}
\label{fig:12}
\end{figure}

To understand the contribution of the SST and the SSS trends on the sea-surface \textit{p}CO\textsubscript2 decadal trend, we perform a sensitivity experiment. In this experiment, we first detrend both the SST and the SSS as depicted in Fig.\ref{fig:12}a. In Fig.\ref{fig:12}, the broken blue lines represent the detrended SSS, whereas the continuous red line represents the detrended SST. Then we feed the trained XGBoost model (a) with the detrended SST and kept the SSS unchanged, and (b) with the detrended SSS and keep the SST unchanged, and predict the sea-surface \textit{p}CO\textsubscript2 for the last decade (Fig.\ref{fig:12}b and c). We observe from Fig.\ref{fig:12}b that as we provide a detrended SST, the value of the negative slope is increases to -0.2005 $\mu$atm per year from -0.4852 $\mu$atm per year. This suggests that the decadal changes in the trend of the \textit{p}CO\textsubscript2 due to the lowering SSS is $\approx$ 41.3\%. Similarly, the contribution of SST increases the slope of the sea-surface \textit{p}CO\textsubscript2 to -0.1842 $\mu$atm per year, which contributes to about $\approx$ 37.9\% of the total negative slope of sea-surface \textit{p}CO\textsubscript2. Hence we find that in the central BoB, the combined contribution of the changes in the SST and the SSS to the trend of sea-surface \textit{p}CO\textsubscript2 is $\approx$ 79.2\% (assume first-order effects are primary), which indicates the effect of other factors such as biology, aerosol deposition, dissolved inorganic carbon, and the total alkalinity etc., has only $\approx$ 30\% responsible for the declining slope of the \textit{p}CO\textsubscript2. The lower contribution from factors apart from SST and SSS in the central BoB region can be attributed to stratification induced thick barrier layer (which inhibits vertical transport and therefore reduces the number of nutrients reaching the sea-surface) \citep{joshi2021influence}, whereas the atmospheric dust is a localized effect; hence may not reach the open oceans of BoB \citep{sridevi2021role}. 

    \begin{table}[ht!]
        \centering
        \begin{tabular}{|c|c|c|c|c|}
        \hline
        \textbf{Variables}&\textbf{Seasons}&\textbf{Mean}  & \textbf{Rate (per year)} & \textbf{P-value} \\
        \hline
        \hline
               & DJF & 27.51 & 0.0138 & 0.5\\
        SST (\textdegree{}C) & MAM & 29.49 & -0.0132 & 0.71\\
               & JJAS & 28.95 & 0.0108 & 0.75\\
               & ON & 29.08 & 0.0348 & 0.13\\
        \hline
        \hline
               & DJF & 32.77 & 0.0045 & 0.72\\
        SSS (PSU) & MAM & 32.70 & -0.0230 & 0.52\\
               & JJAS & 33.02 & -0.0128 & 0.35\\
               & ON & 32.52 & 0.0059 & 0.82\\
        
        \hline
        \hline
        
               & DJF & 378.39 & -0.742 & 0.01\\
        \textit{p}CO\textsubscript2 ($\mu$atm) & MAM & 387.55 & -1.909 & 0.04\\
               & JJAS & 381.85 & -0.208 & 0.54\\
               & ON & 387.72 & 1.212 & 0.04\\
        \hline
        
        \end{tabular}
        \caption{Mean values of SST, SSS, and sea-surface \textit{p}CO\textsubscript2 and the seasonal rate of change of each of these variables.}
        \label{tab:5}
    \end{table}  

To analyze the seasonal variability (over the past decade) in the SST, the SSS, and the sea-surface \textit{p}CO\textsubscript2, we take the seasonal average over each year. We divide seasons according to the Indian monsoon cycle, which is recognized as: The pre-monsoon season (MAM), the southwest monsoon season (JJAS), the post-monsoon season (ON), and the northeast monsoon or the winter monsoon season (DJF). The seasonal variations of each of the variables are displayed in Table \ref{tab:5}. 

The SST is observed to be maximum in the pre-monsoon period and lowest in the winter monsoon season. We notice a seasonal amplitude difference of about 2 \textdegree{}C between the pre-monsoon season and the winter monsoon season. A decrease in SST is perceived over the pre-monsoon season (-0.0132 \textdegree{}C), while it is increasing in all other seasons. The post-monsoon season experiences the highest increment rate (with high significance) in the SST. Interestingly, the SSS is maximum during the southwest monsoon season and lowest during the post-monsoon season. Though the southwest monsoon has the highest river discharge and high precipitation, the freshwater plume begins moving southward in this season. The plume reaches maximum spread in October month \citep{joshi2021influence}, which indicates the freshwater spreads throughout the study domain during the post-monsoon season resulting in minimum SSS. The drop in the amplitude from the southwest to the post-monsoon season is noticed to be by 0.5 PSU. The rate of decrease in the SSS is highest in the pre-monsoon season, followed by southwest monsoon season. The post-monsoon and the winter monsoon experience a low rise in the SSS; hence overall, the SSS is always decreasing. 

From Table \ref{tab:5}, we observe the post-monsoon season to have the highest \textit{p}CO\textsubscript2, but the difference between the pre-monsoon season and the post-monsoon season is $\approx$ 0.2 $\mu$atm, which is almost insignificant. The lowest sea-surface \textit{p}CO\textsubscript2 is seen in the winter monsoon season. The sea-surface \textit{p}CO\textsubscript2 has a significant ($>$ 95\%) rate of decrease in the pre-monsoon season, and it rises almost at a similar rate in the post-monsoon season. Hence the total rate of decrease is due to the lowering sea-surface \textit{p}CO\textsubscript2 trend in the winter and southwest monsoons. This may indicate the effect of increased precipitation on the sea-surface \textit{p}CO\textsubscript2.

\subsection{Summary and Conclusion}
\label{s:3.4}

In this study, we attempt to exhibit the performance of some advanced machine learning algorithms for predicting the sea-surface \textit{p}CO\textsubscript2 using the publicly available raw data. The paucity of the available observation in the BoB has been a hindrance in understanding the dynamics of sea-surface \textit{p}CO\textsubscript2 and in the validation of models. In this study, we collect the data for sea-surface \textit{p}CO\textsubscript2 with the corresponding independent variables from the sources available outside the Exclusive Economic Zone (EEZ) region. We attempt the prediction of sea-surface \textit{p}CO\textsubscript2 based on only two variables (SST and SSS). The \textit{p}CO\textsubscript2 is dependent upon several other factors such as DIC, TALK, biological production, atmospheric dust, and wind, but the data of all the parameters except the SST and the SSS is unavailable along the cruise tracks or the RAMA buoy mooring location. Hence it is difficult to incorporate all the variables upon which \textit{p}CO\textsubscript2 is dependent. To reduce the impact of this limitation on our results, we choose central BoB for our current \textit{p}CO\textsubscript2 predicting endeavour. The central BoB has the least productive zone in the north Indian ocean owing to the presence of high stratification, this central BoB is least impacted by the atmospheric dust because the region is away from the coast. The open oceans are relatively less explored due to the paucity of data giving added importance to this study.

The data visualization shows a low linear relationship between the sea-surface \textit{p}CO\textsubscript2 and the SSS. The low performance of MLR in this study is attributed to this absence of strong linearity. The ANN is a popular choice of model in almost every field, but the XGBoost model has recently gained popularity due to its accuracy in predicting using a low amount of data. We tune the hyper-parameters for both these models (as shown in Tables \ref{tab:3} and \ref{tab:4}) before deploying them. The three models (MLR, ANN, and XGBoost) are compared to the same test data. The Taylor diagram in Fig.\ref{fig:8} reveals the XGBoost model to be the best performing model. We further test the XGBoost model's capability to predict sea-surface \textit{p}CO\textsubscript2 using satellite-derived SST and SSS.

The evaluation of the satellite SST and SSS with the in-situ data from the RAMA buoy shows an excellent agreement between the satellite-derived SST and the in-situ data. Even though the ESA-CCI (satellite merged) SSS product is an improvement over other satellite products, it still performs averagely compared with the RAMA buoy in-situ data (Fig.\ref{fig:9} and Fig.\ref{fig:10}b). The XGBoost predicts the sea-surface \textit{p}CO\textsubscript2 using the satellite-derived products satisfactorily (Fig.\ref{fig:10}c). We compare the performance of XGBoost with some recent studies and found that the XGBoost indeed is better in emulating the sea-surface \textit{p}CO\textsubscript2 for BoB. 

We further investigate the inter-annual variability of the SST, the SSS, and the sea-surface \textit{p}CO\textsubscript2 for the recently passed decade (2010-2019) in the central BoB region (Fig.\ref{fig:11}). The study shows that in the past 10 years, the SST has risen at a rate of 0.0175 \textdegree{}C per year, whereas the SSS is declining at a rate of -0.0088 per year due to an increased precipitation and Himalayan ice melting. The XGBoost predict the sea-surface \textit{p}CO\textsubscript2 to have decreased at a rate of -0.4852 $\mu$atm per year with $\>$ 90\% significance between 2010-2019. The sensitivity experiments (by detrending one variable at a time) reveal that the variability of the SSS from 2010-2019 affects the total declining trend of the sea-surface \textit{p}CO\textsubscript2 by $\approx$ 41\%, and the SST trends affect by $\approx$ 37\% (Fig.\ref{fig:12}). The seasonal variation revealed that in the pre-monsoon season the sea-surface  \textit{p}CO\textsubscript2 has the highest rate of decrease. The lowest sea-surface \textit{p}CO\textsubscript2 is seen in the winter monsoon seasons, and maximum in the post-monsoon season. 

The study reveals the better performance of some advanced machine learning algorithms in predicting the sea-surface \textit{p}CO\textsubscript2 in the BoB. It indicates that if the data inside the EEZ region is also added to this study; these advance algorithms can predict sea-surface \textit{p}CO\textsubscript2 better. Further this approach could be used to predict the effects of the rising tropical cyclones on the sea-surface \textit{p}CO\textsubscript2 in the BoB region. The past decade has shown a decrement in the sea-surface \textit{p}CO\textsubscript2, which shows that the open oceans of the BoB are probably still a sink of the atmospheric CO\textsubscript2. The increasing sink strength will aid in modulating the coastal increase of the sea-surface \textit{p}CO\textsubscript2 due to the influence of anthropogenic activities.

 \section*{Acknowledgment}
 
The authors are grateful to the Indian Institute of Technology. Kharagpur for providing the facilities to conduct the present study. The first and second authors would like to thank the Ministry of Education, Government of India, for the fellowship assistance.






\bibliographystyle{elsarticle-harv} \biboptions{authoryear}
\bibliography{sample.bib}







\end{document}